# Shear-Induced Unfolding Activates von Willebrand Factor A2 Domain for Proteolysis

Carsten Baldauf[a,b], Reinhard Schneppenheim[c], Wolfram Stacklies[a,b], Tobias Obser[c], Antje Pieconka[d], Sonja Schneppenheim[d], Ulrich Budde[d], and Frauke Gräter[a,b,e] a CAS-MPG Partner Institute for Computational Biology, Shanghai Institutes for Biological Sciences, Chinese Academy of Sciences, Shanghai, P.R. China

b BioQuant, Heidelberg University, Heidelberg, Germany

c Department of Pediatric Hematology and Oncology, University Medical Center Hamburg-Eppendorf, Hamburg, Germany

d Coagulation Lab, AescuLabor Hamburg, Hamburg, Germany

e Max-Planck-Institute for Metals Research, Stuttgart, Germany

Corresponding authors:

Carsten Baldauf and Frauke Gräter

CAS-MPG Partner Institute for Computational Biology, Shanghai Institutes for Biological Sciences, Chinese Academy of Sciences, 320 Yue Yang Road, Shanghai, P.R. China

Tel: +86 2154920475; Fax: +86 2154920451

E-mail: carsten@picb.ac.cn, frauke@picb.ac.cn

Reinhard Schneppenheim

Department of Pediatric Hematology and Oncology, University Medical Center Hamburg-Eppendorf, Martinistraße 52, D-20246 Hamburg, Germany

Tel: +49 40428034270; Fax: +49 40428034601

E-mail: schneppenheim@uke.de

**Abstract**

To avoid pathological platelet aggregation by von Willebrand factor (VWF), VWF multimers are regulated in size and reactivity for adhesion by ADAMTS13-mediated proteolysis in a shear flow dependent manner. We examined if tensile stress in VWF under shear flow activates the VWF A2 domain for cleavage by ADAMTS13 using molecular dynamics simulations. We indeed observed stepwise unfolding of A2 and exposure of its deeply buried ADAMTS13 cleavage site. Interestingly, disulfide bonds in the adjacent and highly homologous VWF A1 and A3 domains obstruct their mechanical unfolding. We generated a full length mutant VWF featuring a homologous disulfide bond in A2 (N1493C and C1670S), in an attempt to lock A2 against unfolding. We find this mutant to feature ADAMTS13-resistant behavior *in vitro*. Our results yield molecular-detail evidence for the force-sensoring function of VWF A2, by revealing how tension in VWF due to shear flow selectively exposes the A2 proteolysis site to ADAMTS13 for cleavage while keeping the folded remainder of A2 intact and functional. We find the unconventional 'knotted' Rossman fold of A2 to be the key to this mechanical response, tailored for regulating VWF size and activity. Based on our model we can explain the pathomechanism of some natural mutations in the VWF A2 domain that significantly increase the cleavage by ADAMTS13 without shearing or chemical denaturation, and provide with the cleavage-activated A2 conformation a structural basis for the design of inhibitors for VWF type 2 diseases.

## Introduction

Von Willebrand factor (VWF) is a huge multimeric protein found in blood plasma. VWF mediates the adhesion of platelets to the sub-endothelial connective tissue and is the key protein in primary hemostasis in arterial vessels and the microcirculation (1, 2). Monomeric VWF is synthesized solely in megakaryocytes and endothelial cells. After transfer from the cytosol to the endoplasmatic reticulum, dimers form by C terminal disulfide bonds between CK domains (cf. Fig. 1A). Multimers consisting of up to 100 VWF monomers are then formed in the Golgi and post-Golgi compartment by cystin formation between the N terminal D3 domains. VWF is highly glycosylated, oligosaccharides make up about 20% of the mass of VWF (3). The multimers are either constitutively secreted or stored in endothelial Weibel-Palade bodies and platelet $\alpha$-granules and released from these storage organelles by adequate stimuli. The VWF multimers released from storage are particularly rich in ultra-large VWF (ULVWF). These highly active forms get rapidly yet only partially cleaved by the protease ADAMTS13 at the cleavage site Tyr1605—Met1606 within the A2 domain (4, 5). ADAMTS13 is a zinc containing metallo-protease from the ADAMS/ADAMTS family. Shear stress in blood vessels has been shown to drive VWF multimers into an elongated conformation with increased activity for adsorption to the blood vessel surface, a mechanism to stop bleeding after mechanical injury (6, 7). Mechanical forces due to shear flow regulate selective cleavage of ULVWF and thereby their size distribution (8, 9). If this size regulation fails, ULVWF accumulates and results in phenotypic manifestation of thrombotic thrombocytopenic purpura (TTP) (10). In contrast, reduced VWF concentration or complete absence of VWF results in the different types of von Willebrand disease (VWD) (11), the most common inherited bleeding disorder in humans. While the shear stress induced adhesion and cleavage have been demonstrated in detail *in vitro*, the underlying molecular mechanism of shear induced activation of VWF for ADAMTS13 cleavage is currently unknown.

Structural information at atomic detail on the VWF is scarce. A single VWF is a multi-domain protein featuring a multitude of functionalities (Fig. 1A). Dimerisation and multimerisation are mediated by domains CK and D3, respectively. The central A domain triplet is pivotal for adhesion and clotting, featuring binding sites for collagen (A1, A3) and glycoprotein Ib (GPIb, A3), and the ADAMTS13 cleavage site (A2). A1 and A3 have been shown by X-ray crystallography (12, 13) and A2 by homology modeling to adopt a Rossman $\alpha/\beta$-fold (14). The ADAMTS13 cleavage site in A2 appears to be buried, suggesting that forces in stretched VWF multimers induce unfolding and exposure (15).

We here reveal the unfolding and activation mechanism of A2 for ADAMTS13 cleavage under force by molecular simulations. By applying force distribution analysis, a method previously introduced by our group,(16) we reveal how the atypical Rossman fold topology of the VWF A2 domain senses mechanical force by selectively exposing and activating the ADAMTS13 cleavage site. Furthermore, we predict and analyze the impact of mutations stabilizing the A2 domain by introducing a disulfide bond into VWF A2, in analogy to A1 and A3. We demonstrate this mutant VWF to be resistant against ADAMTS13 *in vitro*. Our results clearly show VWF A2 domain unfolding as a response to shear stress to be the essential event in VWF size regulation.

## Results and Discussion

### *Homology modeling of the VWF A2 domain*

To reveal the molecular process of enforced unfolding and activation of A2 at atomic detail, a homology model including residues 1488 to 1676 of human VWF was created (Fig. 1B). The model fully includes the very terminal sequences of A2, and thereby the site of mutagenesis for introducing a disulfide bond (see below). It is therefore more comprehensive but otherwise highly similar to a previous homology model that covers only the VWF residues 1496 to 1669 (14). The multiple sequence alignment and ProSA 2003 (17) results are shown in the Supporting Information. The model was subjected to equilibrium Molecular Dynamics (MD) simulations. Within the 30 ns simulation time for each of the three independent trajectories the structures converged fast to a backbone root mean square deviation (rmsd) between 0.2 and 0.25 nm (Supporting Information Fig. S3). The agreement with the previous model and the overall high stability indicate the quality of this A2 model and its appropriateness for the subsequent studies.

The secondary structure elements of VWF A2 are organized in the typical Rossman fold as follows (Fig. 1C): β1 L1497 to E1504), α1 (E1511 to Q1526), β2 (I1535 to Y1542), β3 (V1546 to P1551), α2 (D1560 to R1566), α3 (T1578 to D1587), β4 (P1601 to T1608), α4 (R1618 to G1621), β5 (Q1624 to V1630), α5 (Q1635 to R1641), β6 (P1648 to I1651), α6 (F1654 to C1670). The ADAMTS13 cleavage site Y1605—M1606 is located on strand β4 in the protein core, buried on all sides by the surrounding helices and strands, and thus inaccessible for ADAMTS13 in this conformation.

***Force induced unfolding of the A2 domain***

*In vivo*, the size of VWF multimers is regulated by ADAMTS13 in a shear flow dependent manner. The shear flow elongates the VWF and results in a tensile force propagating throughout all VWF domains including A2 in the stretched protein (6, 7). We examined the effect of tensile stress on the VWF A2 conformation by force-probe MD simulations. The effective force in an elongated VWF multimer was accounted for by subjecting the termini of A2 to a pulling force in opposite directions. The resulting force profiles for three independent simulations are shown in Fig. 2. The initial conformation (snapshot 1, Fig. 2) is stepwise unfolded. Starting from the C terminus the secondary structure elements are sequentially peeled-off, namely of α6, β6, and α5 to yield a first intermediate (snapshot 3, Fig. 2), followed by β5 and α4 leading to exposure of the cleavage site (snapshot 4, Fig. 2). Overall, inter-β-strand interactions show higher mechanical resistance than interactions involving helices. A short movie in the Supporting Information illustrates the sequential unfolding of VWF A2 under force.

***Structural characterization and force distribution analysis of the VWF A domain***

The design of VWF A2 apparently is tailored for its force-regulated function. The stable N terminal β1-strand is locked to the center of the protein keeping the protein core including the cleavage site largely intact, while the C terminal structural elements, being more responsive to the external force, are pulled out step by step until the cleavage site is accessible. This distinct response of the two halves of the domain is determined by the underlying topology of the VWF A-type domains. The C terminal part of the A2 domain represents a classical Rossman fold with the characteristic sequential order of the secondary structure elements β4–α4–β5–α5–β6–α6, bridging each strand in the parallel β-sheet alignment with an α-helix. The sequential arrangement results in the stepwise unfolding under force described in the previous section. In contrast, the modified Rossman fold of the N terminal half of the A2 domain prevents sequential unfolding. Here, β-strands are swapped such that β1, the strand directly subjected to the external force, is tightly embedded in the protein core, so as to form rupture-resistant interactions to adjacent strands β2 and β4. We further validated the key role of this particular 'knotted' Rossman topology for the mechanical response of A2 by force distribution analysis (FDA). FDA allows to reveal the distribution of internal strain within a structure subjected to an external pulling force by monitoring changes in pair wise atomic forces, denoted ΔF (see Methods), between a strained and relaxed state of a protein, as described before (16). We determined the strain distribution in an early unfolding intermediate (Fig. 2, snapshot 2) from simulations at constant forces of 150 pN and 16 pN for the strained and relaxed state, respectively, applied to the termini of A2. This intermediate, in which the mechanically labile helix α6 is already unraveled, is sufficiently robust to not show major conformational rearrangements at the moderate external force of 150 pN during 20 independent 30 ns simulations, as required for convergence of internal forces by FDA (Supplementary Fig. S4).

For the early unfolding intermediate, the tensile force mainly propagates through the β sheets of the domain (Fig. 3A), following a direct path between the two termini. Force distributes from the most C terminal strand upstream along the sequence via strands β5 and β4 to the very center of the structure, and therewith to strand β1 that now guides this force out of the domain towards the N terminus. The strand β1 virtually shields the structure formed by the elements β1–α1–β2–β3–α2–α3 from force induced unfolding, as a direct results of the unconventional Rossman fold.

Taking a look at inter side chain forces, this is forces that pairs of side chains exert on each other (see Methods), reveals another interesting feature of the stress propagation in the A2 domain. The cleavage site residue Y1605 is the topological middle point of the domain. Pulling force applied at the termini puts Y1605 under very high strain, mainly created by the side chains of neighboring residues located in the central β strands (Fig. 3A,B, and C). In particular, interactions of Y1605 to V1499, F1501, and V1524 appear to be of importance, as they involve residues located upstream from the cleavage site, and thus remain intact also after further unfolding, i.e. in the intermediate that serves as substrate for ADAMTS13. To test directly, if the A2 unfolding intermediate is selectively stress-activated at the Y1605/M1606 backbone proteolysis site, we performed additional FDA for the cleavage-ready intermediate (Fig. 2, snapshot 4), as above by constant force simulations of the strained and relaxed state, again both stable against unfolding at the time scales of the simulations (Supplementary Fig. S4). Supporting our previous observation, we found high pair wise forces between residues Y1605 and M1606 (Fig. 4D and E). We propose that in addition to mere exposure to ADAMTS13, the Y1605-M1606 proteolytic site in the VWF A2 unfolding intermediate is selectively tensed due to an optimized force distribution, resulting in a weakened peptide bond mechanically activated for cleavage.

**In vitro *mutagenesis and electrophoretic analysis***

Our unfolding simulations suggest A2 to be activated for ADAMTS13 cleavage under high shear flow conditions by exposing the cleavage site after partial unfolding of the C-terminal domain. A1 and A3 have highly similar amino acid sequences and three-dimensional structures, and thus would be expected to unfold along a similar mechanism. Examination of the 3D structures of the VWF A domains shows the existence of disulfide linkages between the termini of the A1 and A3 domains, respectively, but not for the A2 domain. A multiple sequence alignment (Fig. 4A and Supporting Information Fig. S1) shows both A1 and A3 to feature two cysteine residues each at their N and C termini, respectively, allowing the formation of disulfide bridges, rendering A1 and A3 insensitive against tensile stress in VWF under shear. A2 has only two vicinal cysteine residues at its C terminus (Fig. 4B and Supporting Information Fig. S1), which makes this domain mechano-responsive. Thus, the cysteine hooks render A1 and A3 force-resistant and potent for specific interactions with collagen and GPIb as essential for VWF adhesion and aggregation, while allowing the selective force-induced unfolding of only the A2 domain for cleavage by ADAMTS13.

These structural and functional insights on the A domains allow us to design a VWF variant resistant against ADAMTS13. We introduced a cysteine at position N1493 to allow disulfide bond formation with residue C1669 at the A2 C-terminus in analogy to A1 and A3. C1670 was changed to serine to generate maximal homology of A2 with A1 and A3 and to avoid the possibility of alternate disulfide bonding at the A2 carboxy terminal. A model of the mutant A2 domain was subjected to MD simulations to test the feasibility of disulfide bond formation and the domain’s structural integrity upon mutation (Supporting Information Fig. S3B).

To confirm the generation of an artificially introduced A2 disulfide bond *in vitro,* we subjected A2 mutant full length recombinant VWF to multimer analysis in comparison to wild type VWF. Mutant VWF migrated faster than wild type VWF suggesting a more compact structure with higher electrophoretic mobility. In contrast, removing the disulfide bridges in the A1 (C1272S and D1459C) or A3 (C1686S and S1873S) towards an open structure as in wild type A2 resulted in a decrease of the electrophoretic mobility, both in presence (fl) and absence of A2 (delA2), respectively (Fig. 5A and Supporting Information Fig. S5). These results supported our assumption of the generation of a cysteine-bridge connection of the A2 N and C terminus analogous to A1 and A3. We then exposed A2 domain mutant and wild type VWF to ADAMTS13 and monitored proteolysis by multimer analysis. We could show that, in contrast to wild type VWF, ADAMTS13 proteolysis of A2 domain mutant VWF was completely absent, similar as in A2 domain deleted VWF (Fig. 5B). This was further confirmed by reduction of cysteine bonds by β-mercaptoethanol to exclude the possibility that the A2 domain mutant was actually proteolysed but just held together by the created cysteine bonds (Fig. 5C). Opening the disulfide bond of the A1 and A3 domain by mutagenesis in A2 domain deleted VWF did not result in proteolytic susceptibility of the respective domains, indicating that the homology of A1 and A3 to A2 is too low for substrate recognition by ADAMTS13 (Fig. 5B).

**Conclusions**

We here show by simulations and *in vitro* mutagenesis how force-induced partial unfolding is required for ADAMTS13 mediated cleavage of VWF A2. The unfolding and activation mechanism of A2 can be abolished by a single mutation, N1669C, in analogy to the mechanism that presumably protects A1 and A3 from unfolding and loss of function. We find the C-terminal part of VWF A2 to be unraveled under force, suggesting ADAMTS13 to primarily recognize this partially unfolded domain rather than the native state of A2. This is in excellent agreement with recent *in vitro* studies on the interaction of VWF A2 with ADAMTS13 (18, 19). VWF A2 mutations previously identified to cause von Willebrand disease type IIA due to an increased susceptibility to ADAMTS13 (20) cleavage can now on the basis of our model be rationalized. They can be expected to involve destabilization of the overall A2 structure by forcing charged groups into regions of hydrophobic packing (I1628T and G1629E), perturbing β-turn formation between the VWF A2 secondary structure elements β5 and α5 (G1631D), or by destabilizing A2 due to a drastic increase in spatial demand of the side chain (G1609R). Structural destabilization in turn facilitates A2 unfolding and cleavage site exposure to ADAMTS13.

The mechanical unfolding intermediates of the VWF A2 domain observed here and not a static intact equilibrium state from modeling or X-ray crystallography represent the substrate of ADAMTS13. These dynamics of the A2 structure during unfolding are prerequisite to explore the structural and functional determinants of A2 recognition by ADAMTS13 and therewith also to design inhibitors of the enzyme's proteolytic activity, as potential drugs for von Willebrand disease that result from enhanced VWF cleavage in blood.

The force-sensoring mechanism of the A2 domain provides an intriguing explanation for the size regulation of ULVWF: Larger multimers involve higher pulling forces and therefore higher unfolding rates at a given shear flow. As a result, larger VWF is cleaved more readily. The forces required for the exposure of the cleavage site in A2 as observed here (~500pN) can be expected to be significantly larger than those inducing unfolding in *in vivo* conditions due to the short nanosecond time scale of the simulations (21). Under physiological conditions, cleavage shall only occur for the upper limit of VWF multimer sizes, and thus under flow conditions which lead to tensile forces beyond the average 5-10pN estimated for average sizes (6) (A Alexander-Katz, personal communication, 2009).

While the C terminal part of the A2 domain follows a highly conserved unfolding pattern if subjected to tensile stress, the N terminal ‘knotted’ Rossman fold remains completely intact even under high forces. We hypothesize the second important function of A2, the proposed inhibition of the A1 GPIb interaction (22), that mediates the binding of VWF to platelets, to be located at this force resistant part of the domain. Thereby, as a consequence of the two distinct Rossman topologies within the A2 domain, size regulation of VWF by ADAMTS13 does not affect platelet interaction. As a second consequence of the unconventional Rossman fold, we find strain to internally propagate selectively to the ADAMTS13 cleavage site, bringing the peptide bond under tension. We hypothesize the specific force-activation to affect the catalytic function of ADAMTS13, as a direct impact of the A2 mechanics on the A2-ADAMTS13 biochemistry, similar to what has been shown for disulfide bond cleavage by DTT and thioredoxin (23, 24).

We here assumed the stretching force in VWF to propagate to A2 primarily via the covalent inter-domain linkages to adjacent A1 and A3. A full A1-A2-A3 structure is needed to re-examine the unfolding mechanism taking inter-domain interactions into account, as a next important step towards deciphering the molecular details of VWF mechanical response.

Another example for a Rossman fold in which the termini are locked together by a disulfide bond is the VWF type A domain of human capillary morphogenesis protein 2, interestingly again a collagen-binding adhesion protein (25). To what extent nature has made use of the Rossman fold as a module that can be reversibly switched into a force-resistant state remains to be seen.

## Methods

### *Homology modeling and in-silico mutation*

The sequences of the VWF A domains have a residue identity of 20 to 25 %. Based on multiple sequence alignments and structural alignments we created a homology model of the VWF A2 domain (residues 1488 to 1676 of human VWF) and the mutant A2 domain (N1493C and C1670S) from a human VWF A1 X-ray structure (PDB: 1AUQ).

The search for similar sequences was performed in two steps: a fast search with a generalized Fasta methodology and an evaluation based on E-values and Z-scores (26). The structural model comprised residues 1488 to 1676 of human VWF, and therefore a longer sequence than the one used by (14). Finally, pdb-structures (1AUQ, 1ATZ, 1IJB, 1U0O, 2ADF, 1SHU, 1PT6) were selected and subjected to a structural alignment. The resulting sequence alignment is shown in the Supplementary Information. 20 homology models were created from 20 randomized starting configurations, based on the VWF A1 domain (1AUQ). Structures were evaluated on the basis of energies from the Amber99 force field as implemented in MOE. Homology modeling was performed using the molecular operation environment MOE (2007.9, Chemical Computing Group CCG).

Based on the model of the A2 domain, the A2 double mutant N1493C/C1670S was generated. A disulfide bridge was introduced between the termini by the N1493C mutation enabling a link between C1493 and C1669. To maintain a constant content of cysteine residues, known to be beneficial for protein expression (see below), a second mutation C1670S was introduced. The models were validated by molecular dynamics (MD) simulations.

The coordinates of the models are available in PDB format as Supporting Information or upon request from the authors.

### *Molecular dynamics simulation*

All simulations and part of the analysis were carried out with the Gromacs suite of programs (version 3.3.1) (27, 28). The OPLS all atom force field was used for the protein (29, 30). The proteins were solvated in dodecahedric boxes with at least 7,500 TIP4p water molecules (31), and periodic boundary conditions were applied. The typical protonation states at pH 7 were chosen for ionizable groups of the peptide. The necessary amount of counter-ions ($Cl^-$ and $Na^+$) was added to ensure a neutral system. Prior to free MD simulations, steepest descent energy minimizations and position restrained MD simulations with heavy atom positions restrained with a harmonic potential using a spring constant of 1000 $kJ/mol{\cdot}nm^2$ (100 ps) were performed. Temperature (300 K) and pressure (1 bar) were coupled to a Nosé-Hoover thermostat (32, 33) and a Parrinello-Rahman barostat (34, 35), using time constants of 0.1 ps and 1 ps, respectively. Non-bonded interactions were considered within a cut-off of 1 nm, and long-range electrostatic interactions were calculated using the Particle-Mesh-Ewald algorithm (36, 37). Constraints were applied by the Lincs algorithm (38). A time step of 2 fs was used for integration. The wild type and mutant A2 models were simulated three times each for 30 ns and with different seeds for the initial velocity generation. Three independent force-probe MD simulations were performed on a truncated VWF A2 model (residues 1492 to 1670), each ~26 ns in length. Harmonic springs with spring constants of 500 kJ/(mol nm$^2$) were moved away from each other with a velocity of 1.25 nm/ns. To restrict the system size along the pulling direction, after partial unfolding the residues 1636 to 1670 of A2 were removed, water was added to the system, and the force-probe MD simulations were continued. For FDA, a starting system was taken from a snapshot of the unfolding trajectory. Already unfolded parts were removed and a system containing residues 1492 to 1655 was used further on. Constant force of 16 and 150 pN, respectively, for the relaxed and stretched state, was applied in opposing direction to both termini. Each of the two systems was equilibrated under the respective constant force for 20 ns. For both systems, the heavy atom RMSD to the starting structure remained below 0.35 nm for both pulling forces, indicating that the system is able to bear the mechanical stress within this time scale without rupture. In the following, 20 simulations for the folded and 10 simulation for the unfolded state were performed for 30 ns each, starting with different random velocities.

We used the FDA code (16) for Gromacs 4.0 (39) to write out forces $F_{ij}$ between each atom pair $i$ and $j$. Forces were averaged over the total simulation time of 600 and 300 ns per system, respectivley, sufficient to obtain converged averages. Changes in forces, $\Delta F$, were then obtained as the difference in pair wise forces between the systems pulled with 16 and 150 pN. To remove outliers, i.e. some large solvent exposed side chains showing a high $\Delta F$ due to insufficient conformational sampling, we normalized forces with the standard error between individual trajectories as described before (16). Changes in normalized force are denoted $\Delta f$. Residue wise forces $F_{uv}^{res}$ were obtained by summing up forces $F_{ij}$ for all pairs of atoms $i$ and $j$ in residues $u$ and $v$, where atom $i$ and atom $j$ must not be part of the same residue; normalization was done as for inter atomic forces. The absolute sum $\Delta F_u^{res} = \sum_v \left|\Delta f_{uv}^{res}\right|$ reflects the changes in strain acting on a single residue and was used to color code force distribution onto the protein backbone. Strain along the backbone was measured as the sum of all bonded interactions (angles + dihedrals) between adjacent residue pairs. Our MD simulations use LINCS to constrain bond length, and thus no forces for bonds could be calculated. As a result changes in backbone forces indicate strain between two residues, but the values are not physically correct forces. Pymol (http://www.pymol.org) (40), VMD (http://www.ks.uiuc.edu/Research/vmd/) (41) and POV-Ray (http://www.povray.org) were used for visualization.

***VWF engineering and analysis***

By *in vitro* mutagenesis of full length VWF we exchanged N1493 at the N terminal site to cysteine and C1670, one of two neighboring cysteines at the C terminal site of the A2 domain, to serine to allow creation of a cysteine bond in the A2 domain. In additional mutagenesis experiments we also eliminated the existing disulfide bonds in the A1 (C1271S/D1459C) and A3 domains (C1686S/S1873C). *In vitro* mutagenesis of full length VWF cDNA in the mammalian expression vector pcDNA 3.1 by means of the quick change mutagenesis kit (Stratagene) using primers of 41-46 bp in length harboring the particular base exchange, transfer of the cDNA transfection of 293 cells by means of liposomal transfer, cell culture conditions, harvesting and preparing of recombinant VWF and its proteolysis by ADAMTS13 was performed as described previously (20). ADAMTS13-proteolyzed mutant and wild type VWF were also analyzed by polyacrylamide gel electrophoresis under reducing conditions (42).

VWF phenotypic characterization by VWF multimer analysis recorded by digital photo imaging was according to previously published protocols (43-45).

**Acknowledgements**

The authors thank Matthias F. Schneider and Alfredo Alexander-Katz for fruitful discussions. CB thanks René Meier and Wolfgang Sippl (Institute of Pharmaceutical Chemistry, Martin-Luther University of Halle-Wittenberg) for the possibility of a research stay and for the help with the creation of the homology model. CB and FG thank Yandong Yin for performing structure searches. CB is grateful for a Feodor Lynen Fellowship by the Alexander von Humboldt foundation.

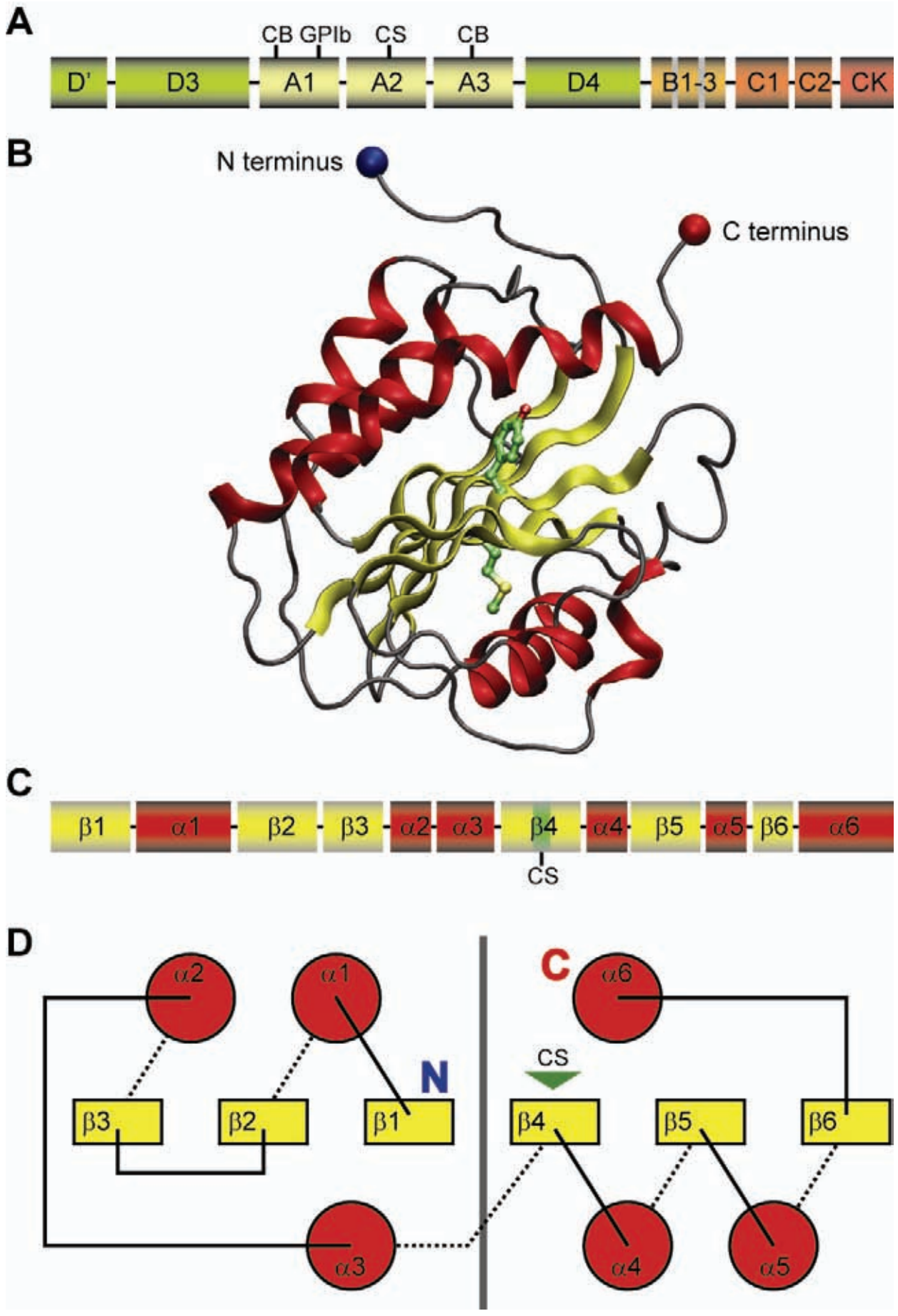

**Fig. 1.** (**A**) Schematic domain organization of the VWF with collagen binding sites (CB) in domains A1 and A3, a glycoprotein Ib (GPIb) binding site in A3, and the ADAMTS13 cleavage site (CS) in A2. (**B**) Homology model of A2 shown as cartoon, the cleavage site is highlighted in green. (**C**) Secondary structure organization, β and α denote β-strand and α-helix, respectively. (**D**) The schematic sketch of the spatial secondary structure orientation shows the classical Rossman-fold of the C terminal half of the A2 domain with CS (green marker) while the N terminal half shows a 'knotted' Rossman fold with significantly higher stability under force.

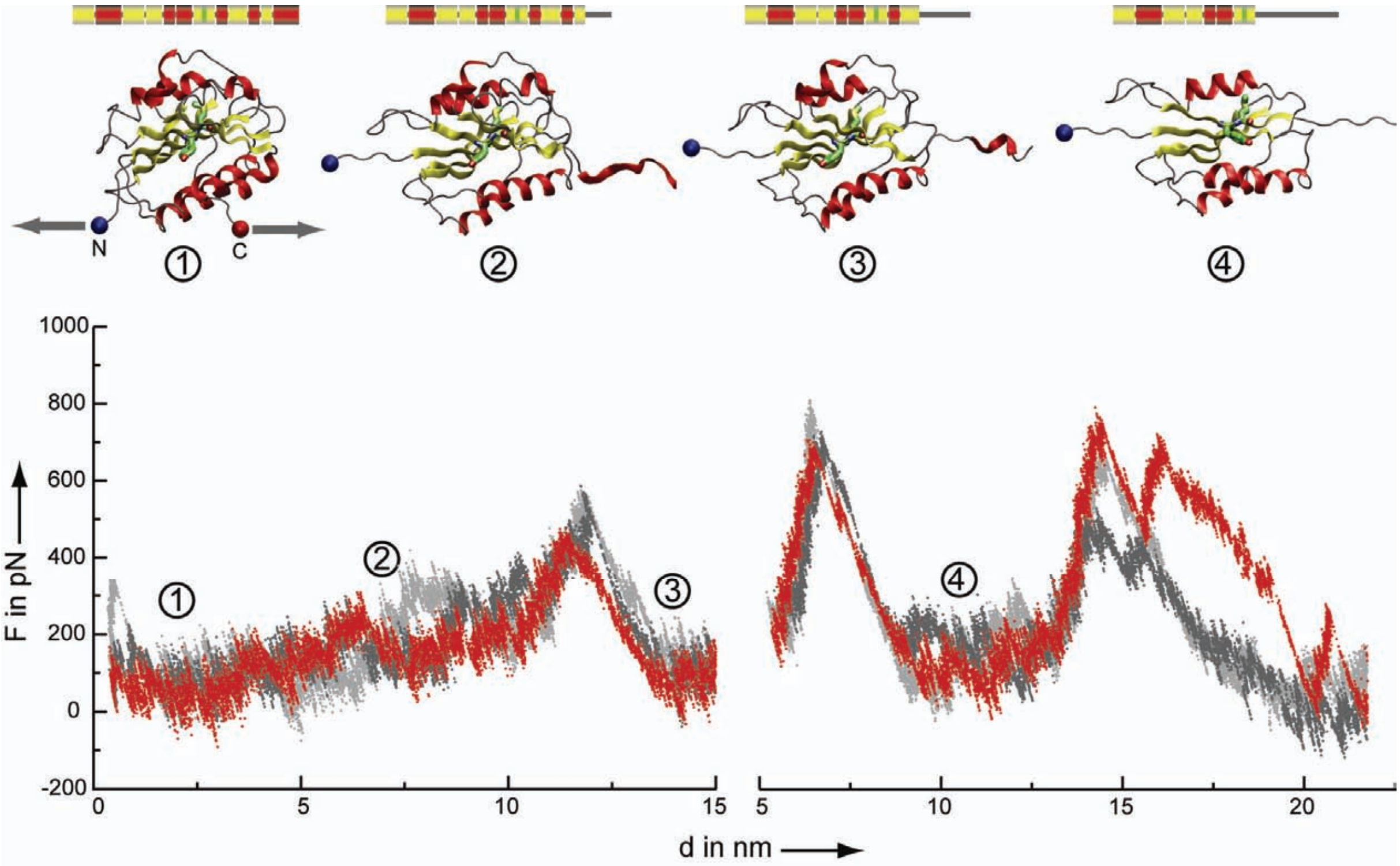

**Fig. 2.** The force profiles for three independent force-probe MD simulations are shown. After extending the protein chain to 15 nm, the simulations were continued with the unfolded C terminal part (sequence numbers 1636 and higher) being cut off. Selected snapshots are shown as cartoon; the cleavage site is shown in green; the fully unfolded C-terminal fragments in 2, 3, and 4 are omitted for clarity.

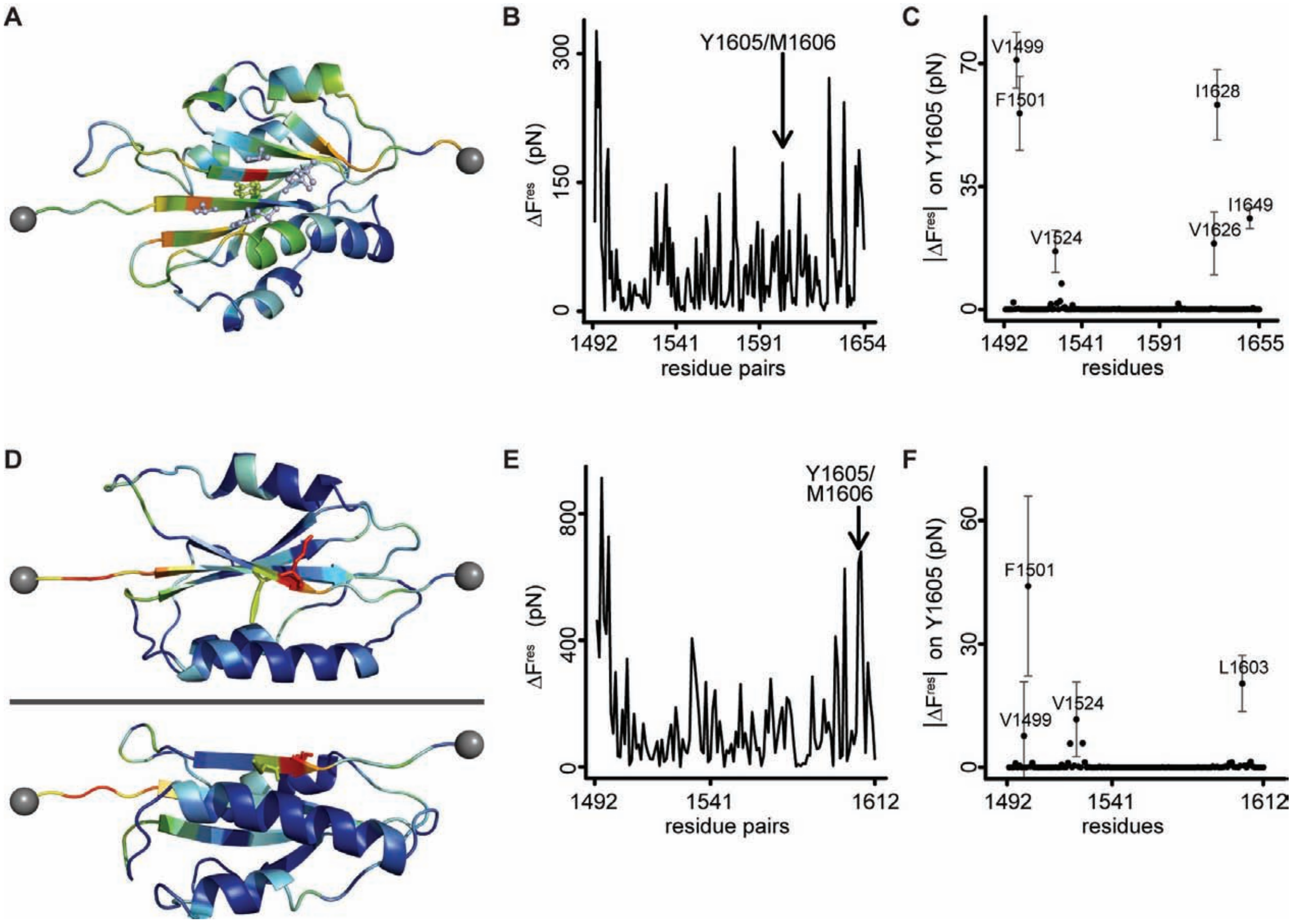

**Fig. 3.** Force distribution analysis (FDA) of a 'folded' state (**A**, cf. snapshot 2 in Fig. 2) and a partially unfolded state (**D**, cf. snapshot 4 in Fig. 2) of the A2 domain: (**A**) Cartoon representation of an A2 folded state. Changes in pair wise forces, Δ*f*, are color coded onto the protein structure. Colors range from blue for Δ*f*=0 to red for high Δ*f*. The external pulling force is mainly distributed along a direct path between the termini, bypassing helices α1 to α4 that are not under strain. Interestingly force propagation involves helix α5, thereby ensuring its early unraveling in the unfolding process. (**B**) Strain along the backbone of the folded structure. The cleavage site turns out to be under high strain, likely supporting cleavage. Strain is measured in terms of changes in bonded interactions between residue pairs. Please note that these values reflect strain at a certain position, but are not physically correct forces (see Methods). (**C**) Strain induced on the cleavage site residue Y1605 by side chain interactions. The plot shows changes in side chain forces $\Delta F^{res}$ for Y1605, standard errors are plotted as whiskers. (**D**) Cartoon representation of the partially unfolded state, colored after backbone strain; color-coding as in **A**. Strain is measured in terms of changes in bonded interactions between residue pairs. Again, we find the cleavage site to be under high strain. Please note that on the C-terminal side, large part of the strain is taken up by non bonded interactions, as reflected by relatively low changes in backbone forces. (**E**) Strain along the backbone of the partially unfolded structure. (**F**) Even in the partially unfolded state, relatively high strain is induced on the cleavage site residue Y1605 by side chain interactions. The plot shows changes in side chain forces $\Delta F^{res}$ for Y1605. In the unfolded state standard errors, plotted as whiskers, are relatively large due to the high flexibility of the now solvent exposed side chains.

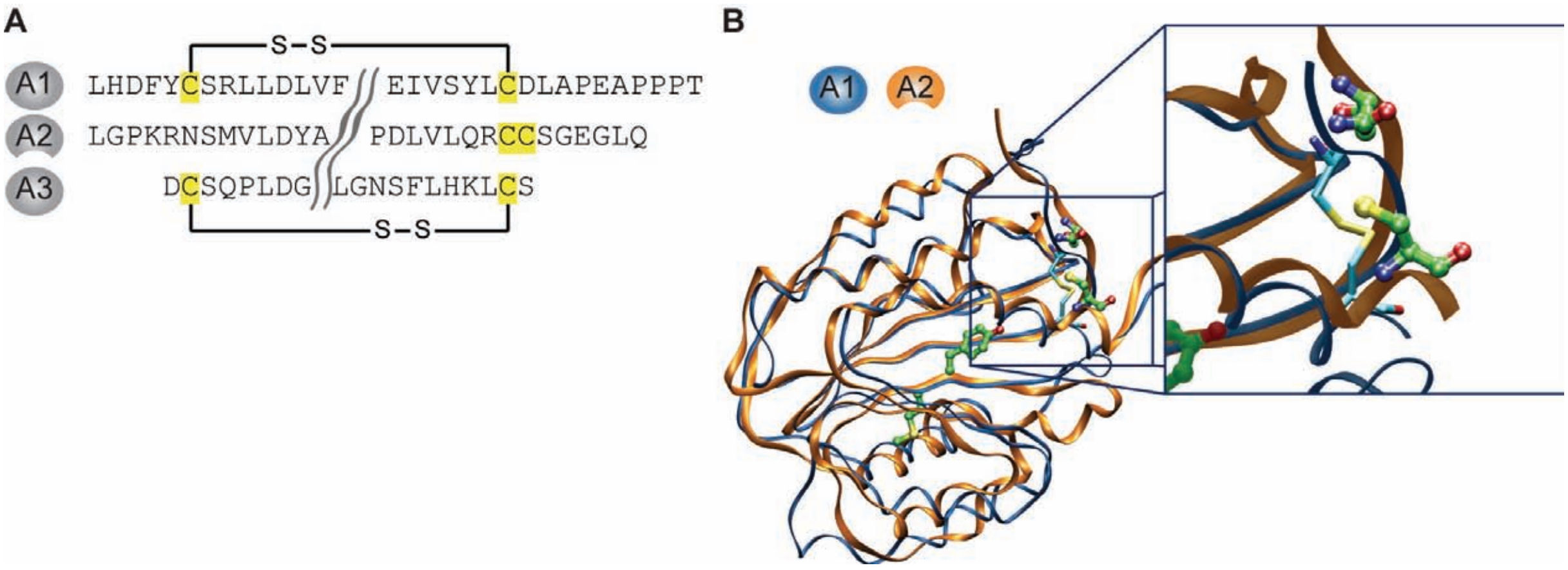

**Fig. 4.** Comparison of the VWF A-type domains: (**A**) Multiple sequence alignment of the A-type domains. Only the N and C-terminal sequences are shown (see Supplementary Fig. S1 for the full alignment), cysteine residues are highlighted and disulfide bonds shown as brackets. (**B**) Structural superposition of A1 and A2. The A1 domain (blue ribbons and blue carbon atoms) is locked by a disulfide bond between the termini, while the A2 domain (orange ribbons and green atoms) lacks this feature. The exchange of residue N1493 for cysteine would allow disulfide bond formation.

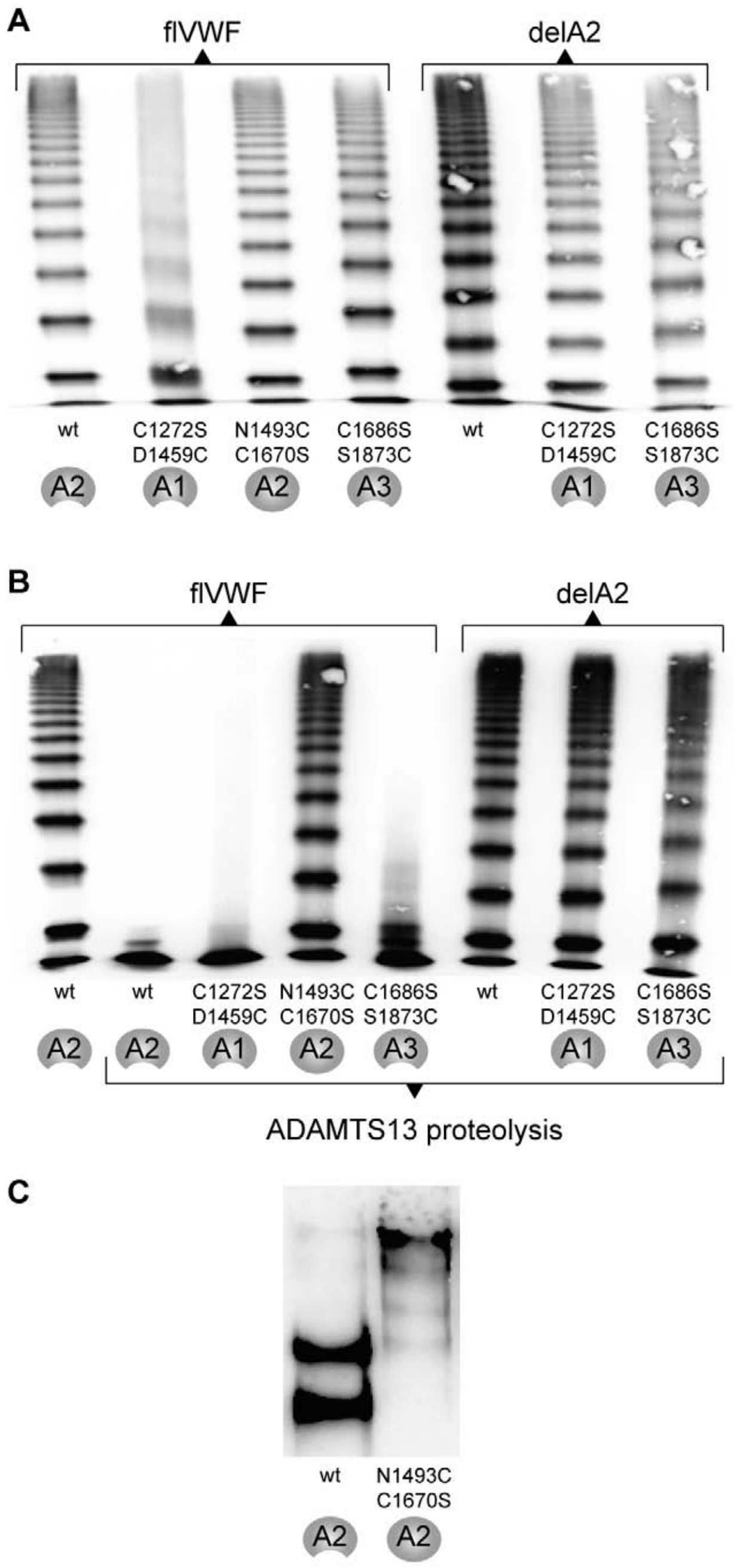

**Fig. 5.** The electrophoretic analysis reveals the protease-stability of delA2 and mutant A2 VWF: (**A**) Multimer analysis of full length (fl) and A2 domain deleted (delA2) mutant and wild type (wt) VWF. Downwards shifted bands indicate faster migration of disulfide bonded mutant A2 (symbol closed circle) and delA2 compared to fl wild type VWF, while mutant A1 and A3 VWF migrate slower due to a more open structure (symbol open circle). (**B**) ADAMTS13 proteolysis of full length (fl) and A2 domain deleted (delA2) mutant and wild type (wt) VWF. Normal proteolysis is seen in wt VWF as well as in A1 and A3 domain mutant VWF with the wild type A2 domain present, whereas proteolysis is absent in A2 domain mutant, and in all delA2 mutants. Open circles denote an open structure of the A domain, closed circles denote disulfide bonded termini. (**C**) Full length A2 domain mutant in comparison to wild type (wt) VWF after ADAMTS13 treatment and reduction of disulfide bonds. Wild type VWF is proteolyzed completely and displays the expected two proteolytic fragments after reduction, whereas mutant VWF is not proteolysed.

**Supporting Information**

# Shear-Induced Unfolding Activates von Willebrand Factor A2 Domain for Proteolysis

Carsten Baldauf[a,b], Reinhard Schneppenheim[c], Wolfram Stacklies[a,b], Tobias Obser[c], Antje Pieconka[d], Sonja Schneppenheim[d], Ulrich Budde[d], and Frauke Gräter[a,b,e] a CAS-MPG Partner Institute for Computational Biology, Shanghai Institutes for Biological Sciences, Chinese Academy of Sciences, Shanghai, P.R. China

b BioQuant, Heidelberg University, Heidelberg, Germany

c Department of Pediatric Hematology and Oncology, University Medical Center Hamburg-Eppendorf, Hamburg, Germany

d Coagulation Lab, AescuLabor Hamburg, Hamburg, Germany

e Max-Planck-Institute for Metals Research, Stuttgart, Germany

Corresponding authors:

Carsten Baldauf and Frauke Gräter

CAS-MPG Partner Institute for Computational Biology, Shanghai Institutes for Biological Sciences, Chinese Academy of Sciences, 320 Yue Yang Road, Shanghai, P.R. China

Tel: +86 2154920475; Fax: +86 2154920451

E-mail: carsten@picb.ac.cn, frauke@picb.ac.cn

Reinhard Schneppenheim

Department of Pediatric Hematology and Oncology, University Medical Center Hamburg-Eppendorf, Martinistraße 52, D-20246 Hamburg, Germany

Tel: +49 40428034270; Fax: +49 40428034601

E-mail: schneppenheim@uke.de

**Fig. S1.** Multiple sequence alignment used as basis for homology modeling.

**1AUQ:** A1 domain of von Willebrand factor (Emsley J, Cruz M, Handin R, Liddington R (1998) J Biol Chem 273: 10396-10401). **1ATZ:** Human von Willebrand factor A3 domain (Huizinga EG, Martijn van der Plas, R, Kroon J, Sixma JJ, Gros P (1997) Structure 5: 1147-1156). **1IJB:** The von Willebrand factor mutant (I546V) A1 domain (Fukuda K, Doggett TA, Bankston LA, Cruz MA, Diacovo TG, Liddington RC (2002) Structure 10: 943-950). **1U0O:** The mouse von Willebrand Factor A1-botrocetin complex (Fukuda K, Doggett T, Laurenzi IJ, Liddington RC, Diacovo TG (2005) Nat Struct Mol Biol 12: 152-159). **2ADF:** Crystal Structure and Paratope Determination of 82D6A3, an Antithrombotic Antibody Directed Against the von Willebrand factor A3-Domain (Staelens S, Hadders MA, Vauterin S, Platteau C, De Maeyer M, van Hoorelbeke K, Huizinga EG, Deckmyn H (2006) J Biol Chem 281: 2225-2231). **2SHU:** Crystal Structure of the von Willebrand factor A domain of human capillary morphogenesis protein 2: an anthrax toxin receptor (Lacy DB, Wigelsworth DJ, Scobie HM, Young JAT, Collier RJ (2004) Proc Natl Acad Sci USA 101: 6367-6372). **1PT6:** I domain from human integrin alpha1-beta1 (Nymalm Y, Puranen JS, Nyholm TKM, Kapyla J, Kidron H, Airenne TT, Heino J, Slotte JP, Johnson MS, Salminen TA (2004) J Biol Chem 279: 7962-7970).

**Fig. S2.** Verification of the Homology Model with ProSA 2003: The energy analysis is smoothed with a window size of 30 aa. Characterizing the model with ProSA-Web shows a Z-score for the raw model of -6.99, and of -8.11 for the model after 10 ns MD simulation. The Z-score for the structure model published by Sutherland et al. is -7.84.

**Fig. S3.** Backbone rmsd of the wild type A2 domain monitored in three independent 30 ns MD simulations. Backbone rmsd of the mutant A2 (N1493C/C1670S) domain monitored in three independent 30 ns MD simulations.

**Fig. S4.** (**A**) Superposition of the average structures under 16 and 150 pN in the folded state used for FDA. Structures are averages over 600 ns, respectively. (**B**) Superposition of the average structures under 16 and 150 pN of the unfolding intermediate. Structures are averages over 300 ns, respectively.

**Fig. S5.** Scan of selected lanes of the gel shown in Figure 4A: The shift of bands relative to the wild type VWF multimers indicates faster migration of mutA2 VWF (N1493C/C1670S) and slower migration of mutA3 VWF (C1686S/S1873).

**Dataset S1.** Homology model of the A2 domain including VWF residues 1488 to 1676 in PDB-format.

**Video S1.** Visualization of a VWF A2 Domain Force Probe MD Simulation:

The N (blue sphere) and C terminus (red sphere) are pulled apart from each other, the secondary structure elements (red: helices; yellow: strands) are stepwise peeled of until the Tyr1605—Met1606 cleavage site (green) is uncovered. A part of the extended and unfolded C terminus is removed in order to save computing time. [xvid4 encoded, AVI container]

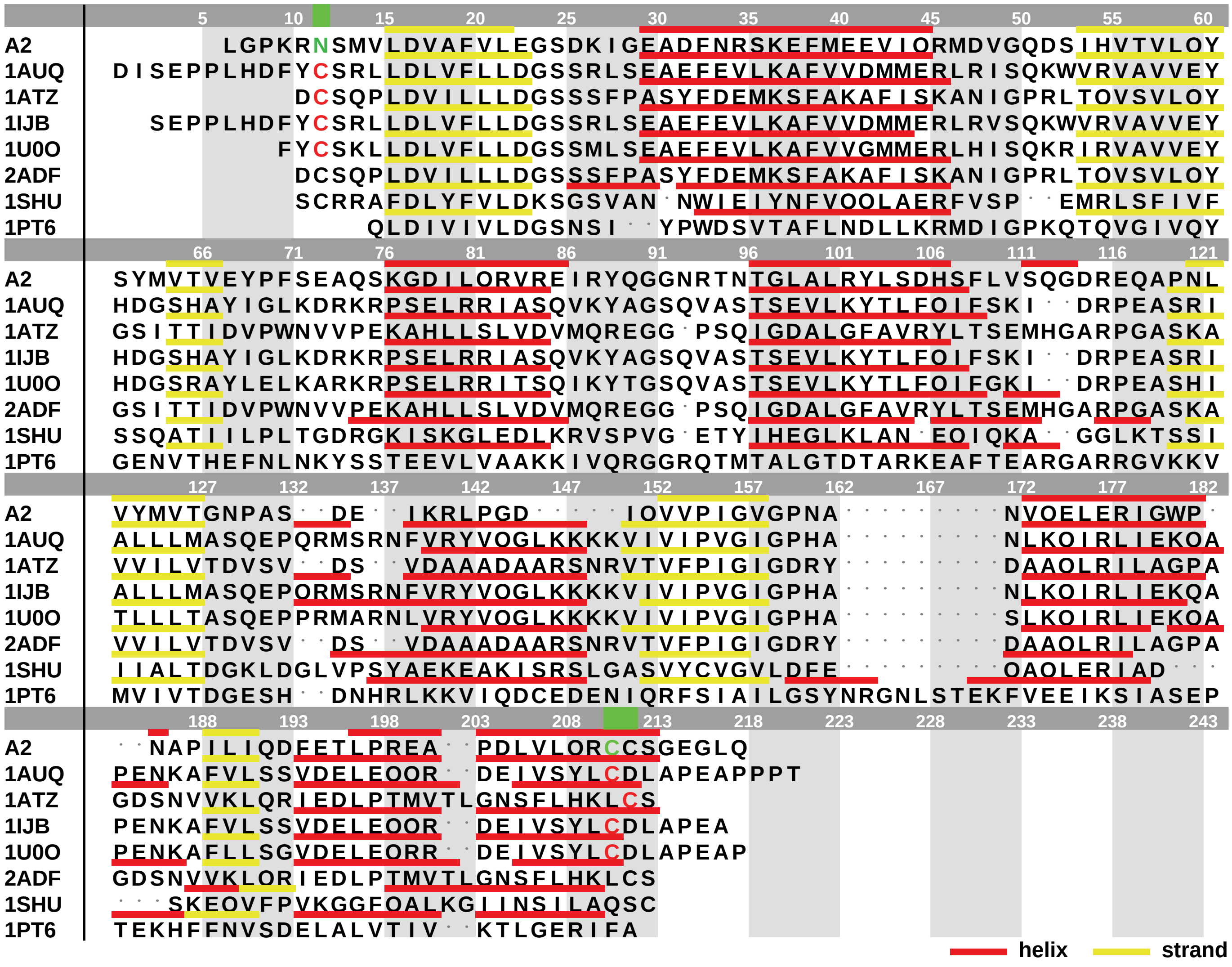

5 10 15 20 25 30 35 40 45 50 55 60
A2 LGPKRNSMVLDVAFVLEGSDKIGEADFNRSKEFMEEVIQRMDVGQDSIHVTVLQY
1AUQ DISEPPLHDFYCSRLLDLVFLLDGSSRLSEAEFEVLKAFVVDMMERLRISQKWVRVAVVEY
1ATZ DCSQPLDVILLLDGSSSFPASYFDEMKSFAKAFISKANIGPRLTQVSVLQY
1IJB SEPPLHDFYCSRLLDLVFLLDGSSRLSEAEFEVLKAFVVDMMERLRVSQKWVRVAVVEY
1U0O FYCSKLLDLVFLLDGSSMLSEAEFEVLKAFVVGMMERLHISQKRIRVAVVEY
2ADF DCSQPLDVILLLDGSSSFPASYFDEMKSFAKAFISKANIGPRLTQVSVLQY
1SHU SCRRAFDLYFVLDKSGSVAN·NWIEIYNFVQQLAERFVSP··EMRLSFIVF
1PT6 QLDIVIVLDGSNSI··YPWDSVTAFLNDLLKRMDIGPKQTQVGIVQY
66 71 76 81 86 91 96 101 106 111 116 121
A2 SYMVTVEYPFSEAQSKGDILQRVREIRYQGGNRTNTGLALRYLSDHSFLVSQGDREQAPNL
1AUQ HDGSHAYIGLKDRKRPSELRRIASQVKYAGSQVASTSEVLKYTLFQIFSKI··DRPEASRI
1ATZ GSITTIDVPWNVVPEKAHLLSLVDVMQREGG·PSQIGDALGFAVRYLTSEMHGARPGASKA
1IJB HDGSHAYIGLKDRKRPSELRRIASQVKYAGSQVASTSEVLKYTLFQIFSKI··DRPEASRI
1U0O HDGSRAYLELKARKRPSELRRITSQIKYTGSQVASTSEVLKYTLFQIFGKI··DRPEASHI
2ADF GSITTIDVPWNVVPEKAHLLSLVDVMQREGG·PSQIGDALGFAVRYLTSEMHGARPGASKA
1SHU SSQATIILPLTGDRGKISKGLEDLKRVSPVG·ETYIHEGLKLAN·EQIQKA··GGLKTSSI
1PT6 GENVTHEFNLNKYSSTEEVLVAAKKIVQRGGRQTMTALGTDTARKEAFTEARGARRGVKKV
127 132 137 142 147 152 157 162 167 172 177 182
A2 VYMVTGNPAS··DE··IKRLPGD····IQVVPIGVGPNA·········NVQELERIGWP·
1AUQ ALLLMASQEPQRMSRNFVRYVQGLKKKKVIVIPVGIGPHA·········NLKQIRLIEKQA
1ATZ VVILVTDVSV··DS··VDAAADAARSNRVTVFPIGIGDRY·········DAAQLRILAGPA
1IJB ALLLMASQEPQRMSRNFVRYVQGLKKKKVIVIPVGIGPHA·········NLKQIRLIEKQA
1U0O TLLLTASQEPPRMARNLVRYVQGLKKKKVIVIPVGIGPHA·········SLKQIRLIEKQA
2ADF VVILVTDVSV··DS··VDAAADAARSNRVTVFPIGIGDRY·········DAAQLRILAGPA
1SHU IIALTDGKLDGLVPSYAEKEAKISRSLGASVYCVGVLDFE·········QAQLERIAD··
1PT6 MVIVTDGESH··DNHRLKKVIQDCEDENIQRFSIAILGSYNRGNLSTEKFVEEIKSIASEP
188 193 198 203 208 213 218 223 228 233 238 243
A2 ··NAPILIQDFETLPREA··PDLVLQRCCSGEGLQ
1AUQ PENKAFVLSSVDELEQQR··DEIVSYLCDLAPEAPPPT
1ATZ GDSNVVKLQRIEDLPTMVTLGNSFLHKLCS
1IJB PENKAFVLSSVDELEQQR··DEIVSYLCDLAPEA
1U0O PENKAFLLSGVDELEQRR··DEIVSYLCDLAPEAP
2ADF GDSNVVKLQRIEDLPTMVTLGNSFLHKLCS
1SHU ···SKEQVFPVKGGFQALKGIINSILAQSC
1PT6 TEKHFFNVSDELALVTIV··KTLGERIFA
helix
strand

**Fig. S1**

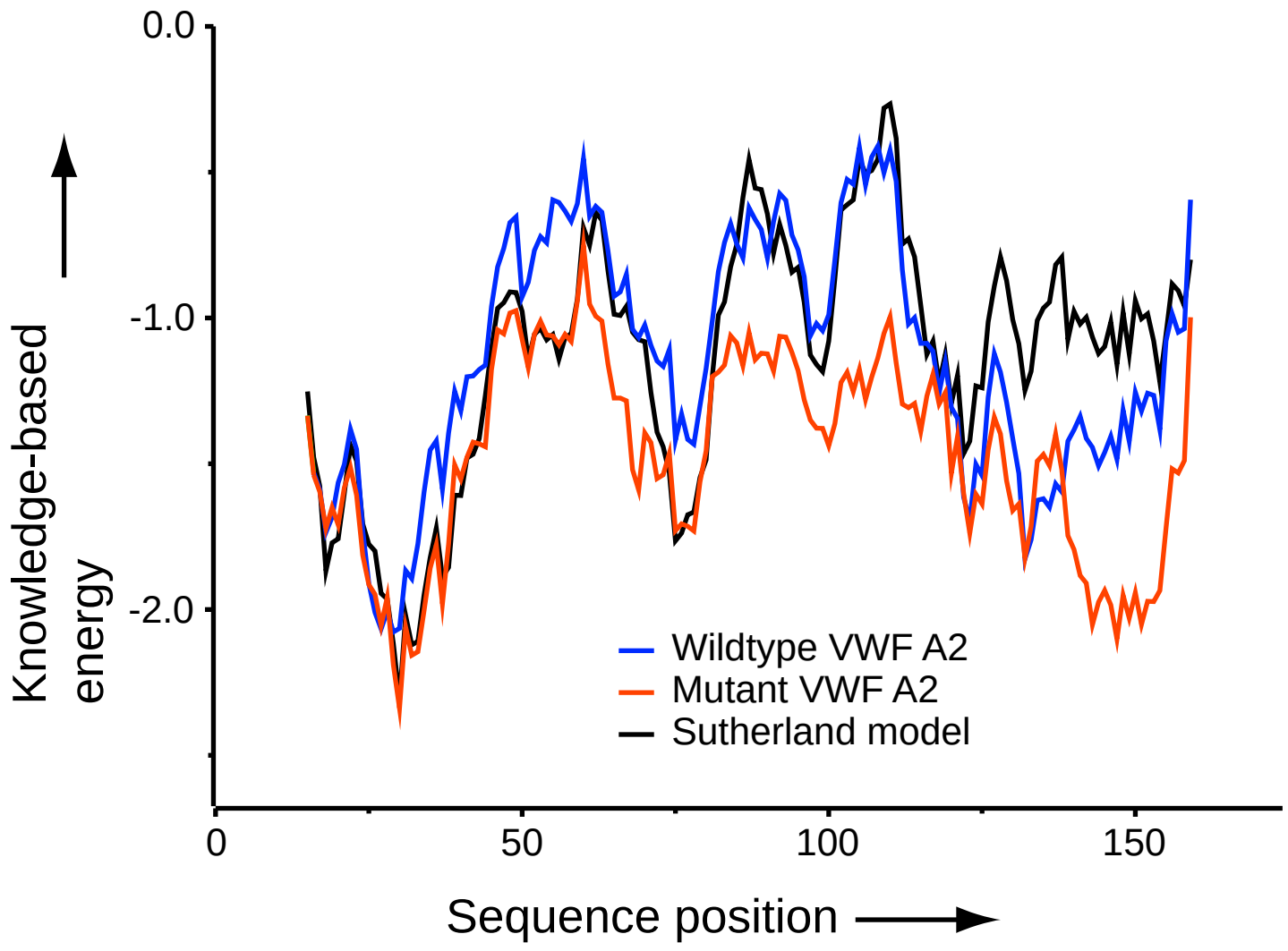

**Fig. S2**

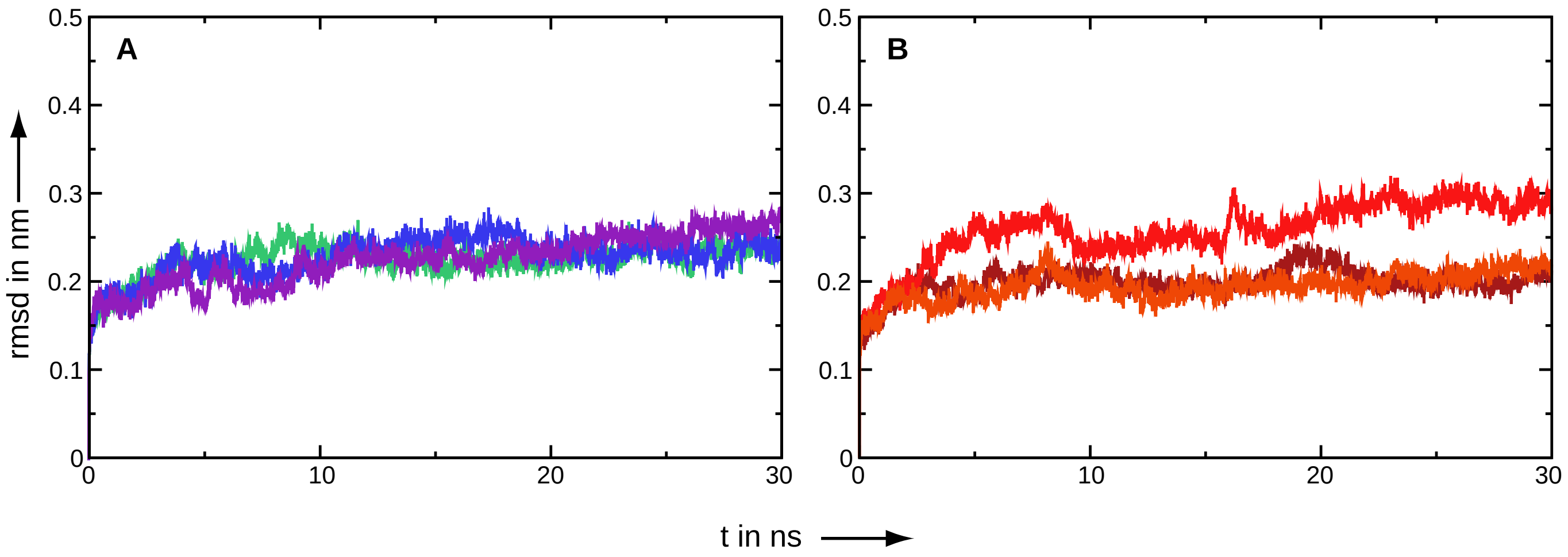

**Fig. S3**

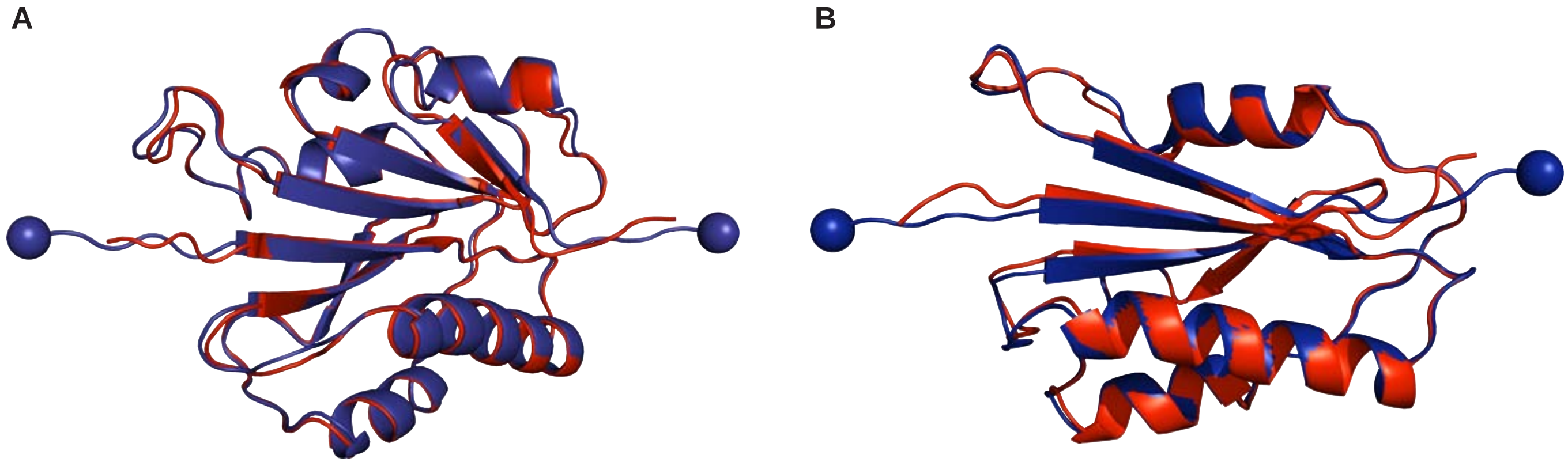

**Fig. S4**

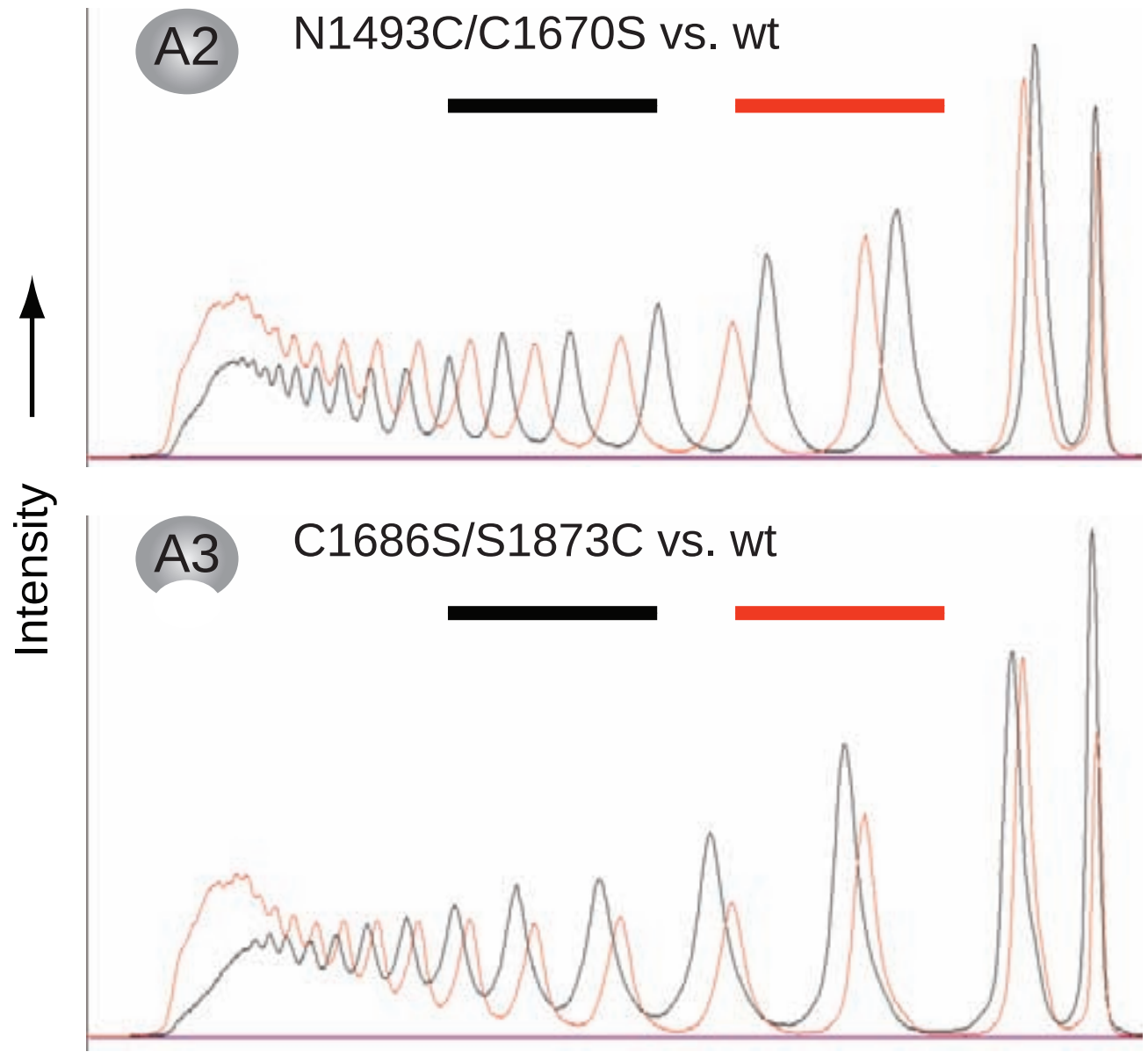

**Fig. S5**